\documentstyle[12pt]{article}
\begin{document}
\baselineskip=14pt

\begin{center}
{\large ESTIMATION OF PROPERTIES OF LOW-LYING EXCITED STATES OF HUBBARD 
MODELS : A MULTI-CONFIGURATIONAL SYMMETRIZED PROJECTOR QUANTUM MONTE CARLO 
APPROACH}
\footnote{Contribution no. 1220 from the Solid State
and Structural Chemistry Unit } \\
\vspace{1cm}
{Bhargavi Srinivasan$^{3}$, S. Ramasesha$^{3,5}$ and H. R.
Krishnamurthy$^{4,5}$  }
{\footnote{e-mail: bhargavi@sscu.iisc.ernet.in,~ramasesh@sscu.iisc.ernet.in,\\
{~~~~~~hrkrish@physics.iisc.ernet.in}} }\\
\vspace{0.5cm}
{\it
$^{3}$Solid State and Structural Chemistry Unit  \\
Indian Institute of Science, Bangalore 560 012, India \\   
\vspace{0.5cm}
$^{4}$Department of Physics  \\
Indian Institute of Science, Bangalore 560 012, India \\   
\vspace{0.5cm} 
$^{5}$Jawaharlal Nehru Centre for Advanced Scientific Research \\
Jakkur Campus, Bangalore 560 064, India \\
}
\end{center}
\vspace{1cm}

\pagebreak
\clearpage

\begin{center}
{\bf ABSTRACT}\\
\end{center}
We present in detail the recently developed multiconfigurational 
symmetrized-projector quantum Monte Carlo (MSPQMC) method for excited states
of the Hubbard model. We describe
the implementation of the Monte Carlo method for a multiconfigurational
trial wavefunction. We give a detailed discussion of 
issues related to the symmetry of the projection procedure which 
validates our Monte Carlo procedure for excited
states. In this context we discuss various averaging
procedures for the Green function and present an analysis of
the errors incurred in these procedures. We study the ground state
energy and correlation functions of the one-dimensional Hubbard model
at half-filling to confirm these analyses. We then study the 
energies and correlation functions of excited states of Hubbard
chains. Hubbard rings away from half-filling are also studied and
the pair binding energies for holes of $4n$ and $4n+2$ systems
are compared with the Bethe  ansatz results of Fye,
Martins and Scalettar. Our study of the two-dimensional Hubbard model
includes the $4 \times 2$ ladder and the $3 \times 4$ 
lattice with periodic boundary conditions.  The $3 \times 4$ lattice is 
non-bipartite and amenable to exact diagonalization studies and is therefore 
a good candidate for checks on the method.  We are able to reproduce accurately
the energies of ground and excited states, 
both at and away from half-filling. We study the properties 
of the $4 \times 2$ Hubbard ladder with bond-alternation as the correlation 
strength and filling are varied. The method reproduces the correlation 
functions accurately. We also examine the severity of sign-problem for
one- and two-dimensional systems.

\section{Introduction}
The study of the Hubbard model for understanding the basic physics
underlying many electronic phenomena in the solid state 
continues to hold centerstage despite its apparent simplicity.
While the model has exact solutions in very special limits \cite
{liebwu,voll}, more
general and experimentally relevant regimes of the Hubbard model
still defy exact solutions. This has led to the development of 
a variety of variational, 
perturbative and nonperturbative numerical many-body techniques.
In recent years, efficient and reliable nonperturbative approaches 
have been developed to study the Hubbard model on finite lattices
over a wide region
of its parameter space. Amongst these, the projector quantum Monte 
Carlo (PQMC) 
method\cite{sorella,imada} and the 
density matrix renormalization group (DMRG)\cite{white} method 
(for 1-D and quasi 1-D systems) have allowed accurate studies of large
Hubbard clusters.  However, these methods have mainly
been limited to obtaining ground state properties. Clearly, properties of
the excited states and excitation gaps of the model
are important for many purposes. Inspite of the importance of the
excited states, there do not exist numerical many-body methods
of sufficient generality to access these states of Hubbard-like models
for large clusters.

The many-body excited states of small clusters can be obtained from
exact diagonalization methods. The usual procedure for obtaining 
excited states in these methods is to exploit the symmetries 
of the system and block-diagonalize the Hamiltonian\cite{sraovb} in a 
convenient basis. The lowest few eigenvalues in each block can then be 
computed using numerical techniques such as the Davidson or modified
Lanczos algorithms\cite{david,mlanc}. 

The block-diagonalization of the Hamiltonian
matrix in the DMRG scheme is non-trivial because of the choice of basis.
While the DMRG method could yield a few low-lying states, low-lying 
excited states of a chosen symmetry were inaccessible from this technique,
until its recent extension to incorporate crucial symmetries of a given
system\cite{srsym}. This has now made it possible to target excited states 
as low-lying states in a subspace of a chosen irreducible 
representation of the symmetry group of a given system.

The PQMC method has exclusively been a ground state technique
for fermionic systems. Furthermore, even the ground state of the  
Hubbard model for arbitrary filling is inaccessible from the PQMC
method when the non-interacting
ground state has an open-shell structure.
Employing a single configuration as a trial state for
the ground state properties of the Hubbard model in such contexts
results in inaccurate estimates of properties\cite{bormann}.

There have been a few attempts to obtain excited state properties
via quantum Monte Carlo approaches. Ceperley and Bernu \cite{ceperley}
introduced a scheme within a Green function
Monte Carlo method for obtaining the rotational-vibrational
spectra of polyatomic molecules for a given potential function. 
This approach is based on constructing a matrix representation of the 
Hamiltonian using approximate functions which under "time" evolution 
have progressively larger projections onto the space of low-lying 
eigenstates. However, this method has not been used in the context
of the Hubbard model.
Takahashi exploited the translational invariance of spin chains
to obtain the des Cloizeaux-Pearson spectrum within a Green function 
Monte Carlo technique\cite{tak}. There has been no generalization of this 
technique to arbitrary symmetries and its use has been restricted 
to spin Hamiltonians. 

In a recent paper we reported a novel multi-configurational 
symmetrized PQMC (MSPQMC) technique\cite{xtdl} which made it
possible, for the first time, to  obtain energies of {\it{excited states}}
of the Hubbard Hamiltonian, within a Monte Carlo scheme. The method was
illustrated for the excitation gaps of half-filled Hubbard chains. 
In this paper, we describe the MSPQMC method in detail including 
the symmetrized sampling procedure for accurate property estimates. 
We apply the
technque to one- and two-dimensional Hubbard models to obtain
properties of ground and excited states at various fillings
and compare these with exact results for small systems. 
We also present a detailed discussion
of the negative-sign problem encountered in the MSPQMC method. We 
summarize our results in the last section. 

\section{The MSPQMC Method}

In this section we describe the multi-configurational symmetrized 
projector quantum Monte Carlo method for ground state properties of
open shell systems and properties of excited states.  In the first
subsection, we give 
a brief description of the conventional PQMC algorithm to make this
paper self-contained. The reader is referred to other papers\cite{sorella,
imada} for more details. We then describe the generalization of this
method to multi-configurational trial states. However, in targetting
an excited state, the validity of the approximations and transformations
carried out in single configurational Monte Carlo procedure need to 
be re-examined. After describing the implementation of the MSPQMC method, 
this issue is dealt with in detail in 
subsection (2.2) and is shown to lead naturally to the idea of 
symmetrized sampling. Furthermore, the errors arising out of the
averaging procedure are examined and the difference between the estimation
of energy and other properties is highlighted. This section is concluded with
a discussion of the negative-sign problem in the MSPQMC method.

\subsection{Implementation of the MSPQMC method}
The single band  Hubbard Hamiltonian $\hat{H}$ for a system of
$N$ sites, may be written  as\cite{hub}, 
\begin{equation}
\hat{H} = \hat{H}_{0} + \hat{H}_{1} = -(\sum\limits_{\langle ij \rangle, \sigma} t_{ij}
\hat{a}^{\dagger}_{i\sigma} \hat{a}_{j\sigma} + h.c.)
+ U\sum_{i=1}^{N}\hat{n}_{i \uparrow}\hat{n}_{i \downarrow},
\label{hamiltonian}
\end{equation}
\noindent          
where the symbols have their usual meanings.
 
Using the projection  ansatz, the lowest eigenstate,
$|\psi_{0}^{\Gamma}\rangle$, in a 
given irreducible symmetry subspace $\Gamma$, of  
$\hat{H}$, can be projected from a trial wavefunction $|\phi^{\Gamma}\rangle$  
as
\begin{equation}
|\psi_{0}^{\Gamma}\rangle
 = \lim\limits_{\beta \rightarrow \infty} { {e^{- \beta{\hat H}}
|\phi^{\Gamma}\rangle} \over {\sqrt{\langle\phi^{\Gamma}| 
e^{- 2 \beta{\hat H}} 
|\phi^{\Gamma}\rangle}}},
\label{projec}
\end{equation}
\noindent
provided $|\phi^{\Gamma}\rangle$ has a nonzero projection on to 
$|\psi_{0}^{\Gamma}\rangle$. This principle was first used in diffusion
Monte Carlo simulations of quantum systems wherein the trial state
is evolved using a random walk algorithm to obtain a stationary 
solution corresponding to the ground state of the system\cite{anderson}.
In the context
of the Hubbard model, however, this ansatz is implemented using
 the Trotter 
formula and the Hubbard-Stratanovich transformation to
estimate expectation values of operators in the ground state
of the Hamiltonian, without explicitly computing the ground state
wavefunction.

In simulations of the Hubbard model, the trial wavefunction 
$|\phi^{\Gamma}\rangle$ is usually formed 
from the molecular orbitals (MOs) obtained
as eigenfunctions of the non-interacting part, $\hat{H}_0$, of the
full Hamiltonian. 
When the non-interacting ground state of a given system is a 
closed-shell state, the trial wavefunction
$| \phi^{\Gamma} \rangle$ for obtaining the interacting ground
state, $|\psi_{0}^{\Gamma}\rangle$, is usually chosen to be a
single nondegenerate electronic configuration in the MO basis. 
Such a choice is adequate to ensure convergence to the ground state for
reasonable values of the projection parameter, $\beta$ and 
Monte Carlo parameters. 

Within the  framework of the projection ansatz, one can clearly target
excited states as the lowest states in various symmetry subspaces,
by choosing trial wavefunctions of the appropriate symmetry.  
However, for this purpose, a single MO-configuration is
no longer an adequate trial wavefunction, since a symmetrized trial
wavefunction,  $| \phi^{\Gamma} \rangle$, is usually 
a symmetrized linear combination of {\em {degenerate}} excited
MO-configurations.
Such a linear combination corresponding to the desired irreducible 
representation, $\Gamma$,
can be obtained, atleast formally, by operating with  
the group theoretic projection operator, $\hat{P}^{\Gamma}$\cite{tinkham}:
\begin{eqnarray}
\hat{P}^{\Gamma} = \sum\limits_{\hat{R}} \chi^{\Gamma}(\hat{R})
\hat{R}
\label{npo}
\end{eqnarray}
\noindent
where  $\chi^{\Gamma}(\hat{R})$
is the character of symmetry element $\hat{R}$ in the $\Gamma^{th}$ 
irreducible representation,
on a single excited MO-configuration. In particular,
to fix the total spin, $S$
of the target state, we use the L\"owdin\cite{lowdin} projection 
operator, $\hat{P}_{S}$,
\begin{eqnarray}
\hat{P}_{S} = \prod\limits_{S^{\prime}\ne S} 
\bigl[ \hat{S}^2 - S^{\prime} (S^{\prime} + 1) \bigr],
\label{lpo}
\end{eqnarray}
\noindent
to project out the desired spin state from a trial configuration.
The projection procedure in eqn. (\ref{projec}) conserves 
the symmetry of the initial state and hence 
projects out the lowest energy state of the interacting 
model of that symmetry subspace from the trial state. The trial 
state $|\phi^{\Gamma} \rangle$ in general takes  the form,
\begin{equation}
|\phi^{\Gamma}\rangle = \sum\limits_{j=1}^{p} c_j^{\Gamma} |
 \phi_{j}^{\Gamma} \rangle ~~;
~~| \phi_{j}^{\Gamma} \rangle = | \phi_{j,\sigma}^{\Gamma} \rangle 
| \phi_{j,-\sigma}^{\Gamma} \rangle,
\label{multicon}
\end{equation}
\noindent
where $p$ is the number of MO-configurations in the
symmetry adapted starting wavefunction. 

A single MO-configuration, $|\phi_{j,\sigma}^{\Gamma}\rangle$,
with $M_{\sigma}$ fermions of spin $\sigma$
can be expressed in second quantized form as,
\begin{equation}
|\phi_{j,\sigma}^{\Gamma}\rangle = \prod\limits_{m=1}^{M_{\sigma}} 
\Bigl( \sum\limits_{i=1}^{N}
({\bf{\Phi}}_{\sigma}^{j\Gamma})_{im} \hat{a}_{i\sigma}^{\dagger} \Bigr) 
|0 \rangle
\label{twf}
\end{equation}
\noindent 
where ${\bf{\Phi}}_{\sigma}^{j\Gamma}$ is an $N \times M_{\sigma}$ 
sub-matrix of  the MO
coefficients whose row index, $i$, labels sites and the column
index,  $m$, labels the  MOs  occupied by  electrons of
spin $\sigma$, in the $j^{th}$ MO-configuration 
in the irreducible representation $\Gamma$. 
For example, a choice of multi-configurational
trial wavefunction for the lowest singlet(S) and triplet(T) states in
the $B^-$-space of the half-filled  Hubbard chain of six sites with
electron-hole and inversion symmetries can be written as
\begin{eqnarray}
\nonumber
|\phi_{S}^{B^{-}}\rangle &=& | \phi_{1}^{B^{-}} \rangle + 
| \phi_{2}^{B^{-}} \rangle ~~;\\
|\phi_{T}^{B^{-}}\rangle &=& | \phi_{1}^{B^{-}} \rangle - 
| \phi_{2}^{B^{-}} \rangle ~~;\\
| \phi_{1}^{B^{-}} \rangle &=& [b^{\dagger}_{3\uparrow}b^{\dagger}_{2\uparrow}
b^{\dagger}_{1\uparrow}] |0 \rangle
[b^{\dagger}_{4\downarrow}b^{\dagger}_{2\downarrow}
b^{\dagger}_{1\downarrow}] |0\rangle \\
| \phi_{2}^{B^{-}} \rangle &=&[b^{\dagger}_{4\uparrow}b^{\dagger}_{2\uparrow}
b^{\dagger}_{1\uparrow}] |0 \rangle
[b^{\dagger}_{3\downarrow}b^{\dagger}_{2\downarrow}
b^{\dagger}_{1\downarrow}] |0\rangle ,
\label{b-twf}
\end{eqnarray}
\noindent
where $b^{\dagger}$'s are the creation  operators for the
MOs. The overlap of any two MO-configurations (eqn. (\ref{twf}))
is given by
\begin{equation}
\langle\phi_{j,\sigma}^{\Gamma}|\phi_{j^{\prime},\sigma}^{\Gamma}\rangle =
det {\Bigl[} ({\bf{\Phi}}_{\sigma}^{j\Gamma})^{T}
({\bf{\Phi}}_{\sigma}^{j^{\prime}\Gamma}) {\Bigr]}.
\label{overlap}
\end{equation}

In the PQMC method for the Hubbard model, the projection 
operator $exp(-\beta \hat{H})$ is Trotter decomposed 
as $(exp(-\Delta \tau \hat{H}))^L$ with $L$ imaginary time slices of 
width $\Delta \tau$ $(\beta = L \times \Delta \tau)$.
This is followed by
a discrete Hubbard-Stratanovich (H-S) transformation\cite{temp} of the  
on-site interaction Hamiltonian. The discrete H-S
transformation applied to a single interaction term,
 $exp(\Delta\tau U \hat{n}_{\uparrow} \hat{n}_{\downarrow})$, yields
\begin{eqnarray}
e^{\Delta\tau U \hat{n}_{\uparrow} \hat{n}_{\downarrow} } =
\sum\limits_{s=\pm 1} e^{\lambda s 
(\hat{n}_{\uparrow} - \hat{n}_{\downarrow} )
 - {\Delta\tau U \over 2}(\hat{n}_{\uparrow} + \hat{n}_{\downarrow} )
}
\end{eqnarray}
\noindent
where $s$ is a single H-S field and $\lambda$~=~$2 arctanh
\sqrt{tanh(\Delta\tau U/4)}$ is the H-S parameter.
Thus,  at a given time-slice, $l$, the interaction term can be expressed
as the exponential of a non-interacting Hamiltonian in
terms of Ising-like fields, $s_{il}$, 
\begin{equation}
e^{-\Delta \tau \hat{H}} ~~=~~ 
\sum\limits_{\{{\bf{s}}_{l}\}}
{\hat{X}}_{\sigma}(l,{\bf{s}}_{l}){\hat{X}}_{-\sigma}(l,{\bf{s}}_{l})
\label{expdelta}
\end{equation}
\begin{equation}
{\hat{X}}_{\sigma}(l,{\bf{s}}_{l})=
exp[{{-\Delta \tau}\over {2}} \hat{H_0}] 
\Biggl(
\sum\limits_{\{{\bf{s}}_{l}\}}
exp[\zeta_{\sigma} \lambda\sum\limits_{i} s_{il}\hat{n}_{i\sigma}
 - {\Delta\tau U \over 2} ]
\Biggr)
exp[{{-\Delta \tau}\over {2}} \hat{H_0} ]
\label{xop}
\end{equation}
\noindent
where the summation is over all possible $N$-vectors
${\bf{s}}_{l}$ whose
$i^{th}$ components correspond to the H-S field and
$s_{il}$, $\zeta_{\sigma}$  is +1 (-1) for electrons with $\uparrow$ 
$(\downarrow)$ spin.
Thus, 
\begin{eqnarray}
e^{-\beta \hat{H}}&=&
\sum\limits_{\{s\}}
%{\bf{s}_{L}},{\bf{s}_{L-1}},\ldots,{\bf{s}_{1}}\rightarrow
%\{s\}}
%\sum\limits_{\{s\}} 
\hat{W}_{\sigma}(\{s\})\hat{W}_{-\sigma}(\{s\})~~=~~
\sum\limits_{\{s\}} \hat{W}(\{s\}) \\
\label{expbeta}
\hat{W}_{\sigma}(\{s\})&=&
{\hat{X}}_{\sigma}(L,{\bf{s}}_{L})\ldots
%{\hat{X}}_{\sigma}(l,{\bf{s}}_{l})\ldots
{\hat{X}}_{\sigma}(1,{\bf{s}}_{1} )  
\label{wop}
\end{eqnarray}
\noindent
The action of each term in the 
summation in eqn. (\ref{expdelta}) on a single configuration
$j$ of the trial state of the form in eqn. (\ref{twf})
can be obtained as the left multiplication of the 
$N \times M_{\sigma}$ matrix $\bf{\Phi}^{j\Gamma}_{\sigma}$
by an $N \times N$ matrix,  ${\bf{B}}_{\sigma}(l,{\bf{s}}_{l})$,
given by
\begin{equation}
{\bf{B}}_{\sigma}(l,{\bf{s}}_{l}) = {\bf{b}}_{0}  {\bf{b}}_{1 \sigma}
(l,{\bf{s}}_{l})  {\bf{b}}_{0}
\label{bmatrix}
\end{equation}
\noindent
The matrix $ {\bf{b}}_{0}$ is given by $ exp [ - {\bf{K}}]$, with
$K_{ij} =   - {\Delta\tau\over2} t_{ij} $. 
The matrix  ${\bf{b}}_{1 \sigma}(l,{\bf{s}}_{l})$ is diagonal with
elements $ \delta_{ij} {1\over2} 
  exp [\zeta_{\sigma} \lambda s_{il} - {\Delta\tau U \over 2} ]$.

The expectation value of an operator $\hat{O}$ 
in the targetted
state is given by
\begin{eqnarray}
 \langle \hat{O} \rangle =  {{\langle\psi^{\Gamma}| \hat{O}
 |\psi^{\Gamma}\rangle}
\over
{ \langle\psi^{\Gamma} | \psi^{\Gamma}\rangle}}.
\label{Oexpec}
\end{eqnarray}
\noindent
To compute such expectation values for a {\it single-configurational}
trial wavefunction, we define right and left 
projected states
$ |R^{\Gamma}(l,\{s_{R}\}) \rangle$ and 
$\langle L^{\Gamma}(l,\{s_L\}) |$.
The former
is obtained by projecting the trial wavefunction through the
right Ising lattice $\{s_R\}$ formed by time-slices 1 through $l$, while
the latter is obtained by projecting its transpose
through the left Ising lattice $\{s_L\}$ formed by time-slices $L$ through 
$l+1$,
\begin{eqnarray}
\nonumber
 |R^{\Gamma}(l,\{s_{R}\}) \rangle
~~&=&~~  {\hat{X}}_{\sigma}(l,{\bf{s}}_{l})\ldots
{\hat{X}}_{\sigma}(1,{\bf{s}}_{1})
|\phi^{\Gamma} \rangle\\
 \langle L^{\Gamma}(l,\{s_{L}\}) |
~&=&~ \langle \phi^{\Gamma}| 
{\hat{X}}_{\sigma}(L,{\bf{s}}_{L})\ldots
{\hat{X}}_{\sigma}(l+1,{\bf{s}}_{l}) 
\label{rlket}
\end{eqnarray}
\noindent
The matrices ${\bf{R}}^{\Gamma}(l,\{s_{R}\})$ and 
${\bf{L}}^{\Gamma}(l,\{s_{L}\})$ which generate the states
$ |R^{\Gamma}(l,\{s_{R}\}) \rangle$ and 
$\langle L^{\Gamma}(l,\{s_L\}) |$ 
in a manner analogous to eqn. (\ref{twf})
are given by
\begin{eqnarray}
\nonumber
{\bf{R}}^{\Gamma}(l,\{s_{R}\})  &=& 
%\sum\limits_{j} c_{j}^{\Gamma}
{\bf{B}}_{\sigma}(l,{\bf{s}}_{l})\ldots {\bf{B}}_{\sigma}(1,{\bf{s}}_{1}) 
{\bf{\Phi}}_{\sigma}^{j\Gamma}  \\
{\bf{L}}^{\Gamma}(l,\{s_{L}\})  &=&
%\sum\limits_{j} c_{j}^{\Gamma} 
\Bigl({\bf{\Phi}}_{\sigma}^{j\Gamma}  \Bigr)^{T}
{\bf{B}}_{\sigma}(L,{\bf{s}}_{L})\ldots {\bf{B}}_{\sigma}(l+1,
{\bf{s}}_{l+1}) 
\label{rlmat}
\end{eqnarray}
\noindent
For $l \approx L/2$, the targetted state $|\psi^{\Gamma} \rangle$ 
can be approximated as
\begin{eqnarray}
\nonumber
|\psi^{\Gamma} \rangle \approx
\sum\limits_{\{s_{R}\}} |R^{\Gamma}(l,\{s_{R}\}) \rangle \\
%~~=~~ \sum\limits_{\{s_{R}\}} \hat{W}(\{s_R\}) |\phi^{\Gamma} \rangle\\
\langle \psi^{\Gamma}| \approx 
\sum\limits_{\{s_{L}\}} \langle L^{\Gamma}(l,\{s_{L}\}) |
%~=~ \sum\limits_{\{s_{L}\}}\langle \phi^{\Gamma}| \hat{W}(\{s_L\})
\label{psiapprox}
\end{eqnarray}
\noindent
This allows us to express $\langle \hat{O}\rangle$ as a weighted average
over Ising configurations $\{s\}$, 
\begin{eqnarray}
\langle \hat{O}(l)\rangle = 
\sum\limits_{\{s\}}\omega^{\Gamma}(\{s\}) O^{\Gamma}(\{s\})
\label{oweight}
\end{eqnarray}
\noindent
where the weight $\omega^{\Gamma}(\{s\})$ is given by,
\begin{eqnarray}
\omega^{\Gamma}(\{s\}) =
%{{  \langle \phi^{\Gamma} | \hat{W}(\{s\}) |\phi^{\Gamma} \rangle }
{{  \langle L^{\Gamma}(l,\{s_L\}) | R^{\Gamma}(l,\{s_R\}) \rangle }
\over
{\sum\limits_{\{s\}} \langle \phi^{\Gamma} | \hat{W}(\{s\}) |
\phi^{\Gamma} \rangle }}\\
\label{omega}
\end{eqnarray}
\noindent
and $O^{\Gamma}(l,\{s\})$, the term in the expectation value 
corresponding to the Ising-configuration $\{s\}$is given
by, 
\begin{eqnarray}
O^{\Gamma}(l,\{s\}) = 
{{  \langle L^{\Gamma}(l,\{s_L\}) | \hat{O} |R^{\Gamma}(l,\{s_R\}) \rangle }
\over
{  \langle L^{\Gamma}(l,\{s_L\}) | R^{\Gamma}(l,\{s_R\}) \rangle }}.
\label{oexpec}
\end{eqnarray}
\noindent
For example, we could estimate the equal time Green function, at the
time-slice $l$,
${\bf{G}}_{\sigma}(l,\{s\})$, which we regard as a matrix in the Wannier basis,
for spin $\sigma$, for an Ising-configuration 
$\{s\}$. The $(m,n)^{th}$ matrix element of
${\bf{G}}_{\sigma}(l,\{s\})$  is given by
\begin{equation}
 ({\bf{G}}_{\sigma}(l,\{s\}))_{mn}  
%\langle \hat{a}_{m \sigma} \hat{a}_{n\sigma}^{\dagger}\rangle
= 
{ \langle  L_{\sigma}(l,\{s\})|  \hat{a}_{m \sigma}
\hat{a}_{n \sigma}^{\dagger} | R_{\sigma}(l,\{s\}) \rangle \over
\langle  L_{\sigma}(l,\{s\}) | R_{\sigma}(l,\{s\}) \rangle }
\end{equation}
\noindent
Thus, when the operator $\hat{O}$ is
the single-particle operator, 
$\hat{a}_{m\sigma} \hat{a}^{\dagger}_{n\sigma}$
and the trial wavefunction is a single
MO-configuration, $O^{\Gamma}(l,\{s\})$
can be shown to take the form\cite{imada},
\begin{equation}
%\langle \hat{a}^{\dagger}_{k\sigma} 
%\hat{a}_{l\sigma}\rangle^{\Gamma}(\{s\})~~=~~
O^{l,\Gamma}(l,\{s\}) ~~=~~ ({\bf{G}}_{\sigma}(l,\{s\}))_{mn} ~~=~~
{\Bigl(}  {\bf{R}}^{\Gamma}_{\sigma}(l,\{s\}) 
( {\bf{L}}^{\Gamma}_{\sigma}(l,\{s\}) {\bf{R}}^{\Gamma}_{\sigma}(l,\{s\} )^{-1}
 {\bf{L}}^{\Gamma}_{\sigma}(l,\{s\})  {\Bigr)}_{mn}.
\label{rlrinvl}
\end{equation}
\noindent
%which is the $kl^{th}$ element of the
%single-particle Green function,~~$G^{\Gamma}_{\sigma}(l,\{s\})$\cite{imada}.
If we weight average the property over all the Ising-configurations,
we would obtain the expectation value of that property in the 
targetted state, exact to within Trotter error. 
However, exhausting all Ising-configurations
in an averaging procedure is impractical and the denominator in
eqn. (\ref{omega})  cannot be known explicitly. Therefore,
we resort to an importance sampling Monte Carlo (MC) estimation
in which a knowledge of the ratio of weights, $r$, for any two 
configurations $\{s^{\prime}\}$ and $\{s\}$,
$\omega^{\Gamma}(\{s^{\prime}\})/\omega^{\Gamma}(\{s\})$ is sufficient 
for obtaining property estimates. 
%The ratio, $r$  is given by the ratio of inner products 
%of the right and left projected states  for the two
%Ising-configurations. 
Using eqn. (\ref{overlap}), this ratio
can be written as
\begin{eqnarray}
r =  \prod\limits_{\sigma} r_{\sigma} =
\prod\limits_{\sigma} { {det
 {\Bigl(} {\bf{L}}^{\Gamma}_{\sigma}(l,\{s^{\prime}_{L}\})
{\bf{R}}^{\Gamma}_{\sigma}(l,\{s^{\prime}_{R}\}) {\Bigr)} }
\over
{det {\Bigl(} {\bf{L}}^{\Gamma}_{\sigma}(l,\{s_{L}\})
{\bf{R}}^{\Gamma}_{\sigma}(l,\{s_{R}\}) {\Bigr)} }  } 
\label{rat}
%&=&\prod\limits_{\sigma}{ {det{\bf{g}_{\sigma}}(\{s^{\prime}\})} \over 
%{det{\bf{g_{\sigma}}}(\{s\})} }
%\label{ratg}
\end{eqnarray}

Obtaining the ratio of determinants and the Green function,
in practice, involves obtaining the inverse of a matrix
which is an $O(N^3)$ operation for an $N \times N$  matrix. If a single spin
flip mechanism is used, the ratio for flipping spin $s_{il}$ in the
configuration $\{s\}$ to 
$s_{il}^{\prime}$ obtain the configuration $\{s^{\prime}\}$,
$r$, can be expressed as
\begin{equation}
r = \prod\limits_{\sigma} { det ( {\bf{L}}_{\sigma}^{\Gamma}
(l,\{s_{L}^{\prime}\}) 
(1 + {\bf{\Delta}}_{\sigma}) 
R_{\sigma}^{\Gamma}(l,\{s_{R}^{\prime}\}) ) \over
det( {\bf{L}}_{\sigma}(l)  {\bf{R}}_{\sigma}(l) ) }
\label{rdelta}
\end{equation}
\noindent
where
\begin{equation}
{\bf{ \Delta}}_{\sigma j j^{\prime}} = \cases{  \delta_{\sigma} = exp (\lambda
\zeta_{\sigma}(s_{il}^{\prime} - s_{il})) -1 & for $j = j^{\prime} =i $
\cr
%     -\delta_{\downarrow} = exp (-\lambda
%(\sigma_{i}^{\prime}(l) - \sigma_{i}(l))) -1 & for $j = j^{\prime} =1 
%and \sigma = \downarrow$  \cr
                           0 & otherwise}         
\end{equation}
\noindent
The ratio of determinants $r_{\sigma}$ can be written in terms of the 
Green function as,
\begin{equation}
r_{\sigma} = | 1 + \delta_{\sigma} ( {\bf{G}}_{\sigma}(l,\{s\}) )_{ii}|
\end{equation}
\noindent
Once the new configuration $\{s^{\prime}\}$ is accepted, its Green 
function $G_{\sigma}^{\prime}(l)$ (dropping arguments for clarity)
can be calculated from the Green function of the old configuration,
$G_{\sigma}(l,\{s\})$ as
\begin{equation}
({\bf{G}}_{\sigma}^{\prime}(l))_{mn} =  
({\bf{\Delta}}_{\sigma}{\bf{b}}_{1})_{mm} ({\bf{b}}_{1}^{-1})_{mm} 
\Bigl[ ({\bf{G}}_{\sigma}(l))_{mn} - { ({\bf{G}}_{\sigma}(l))_{mi} 
\delta_{\sigma} 
({\bf{G}}_{\sigma}(l))_{in} \over 
1 + ({\bf{G}}_{\sigma}(l))_{ii}\delta_{\sigma}} \Bigr]
\end{equation}
\noindent
In the ususal PQMC procedure, Ising-configurations are 
generated by sequential single spin-flips through the lattice, 
examining each site at a given time slice, $l$, before 
proceeding to the next. The matrix elements of the 
Green function are computed at the
time slice at which Ising spin flips are attempted.
This allows the use of the $O(N^2)$ updating 
algorithm described above for the Green function instead of the 
$O(N^3)$ direct algorithm.
Using the heat bath algorithm, the new configuration
is accepted or rejected with a probability $r/(1+r)$.
However, since the Green function is  obtained through an updating
procedure, it starts to degrade numerically
as the number of spin flips increases. Therefore, at suitable intervals, 
we recompute the Green function, using eqn. (\ref{rlrinvl}). We use
the modified Gram-Schmidt orthogonalization procedure of Imada and
Hatsugai\cite{imada} to orthogonalize the columns (rows) of the
right (left) projected trial wavefunction every few "time"-steps
(usually 10). 
The use of an $O(N^{2})$ updating algorithm forces estimation of
the Green function at the time-slice at which a spin flip is attempted.
Therefore, this algorithm yields  properties as averages over all time-
slices. However, the use of a direct $O(N^{3})$ algorithm allows the
possibility of estimating properties at any fixed time, independent of
the time-slice at which the spin-flip is attempted.
In  such calculations, properties could be computed variously by choosing a
particular time-slice $l$ and obtaining all quantities at this 
time-slice, or by averaging over time-slices as in the procedure
described above. We have implemented these methods and demonstrate in
the next subsection that they have a direct
bearing on the symmetry of the projection procedure and also discuss their
relative merits.

We now discuss the extension of this method to a multiconfigurational
trial wavefunction.  In its implementation, the MSPQMC procedure is
a generalization of the single configurational procedure and proceeds
identically for ground states of open-shell systems and for excited 
states. However, at a formal level, issues related to conserving
the symmetry of the trial state, essential for targetting excited states
need to be clearly examined, in view of the randomness introduced
by a Monte Carlo sampling procedure. These concerns will also be addressed in
the next subsection.

Using the PQMC formalism, the Monte Carlo for open shells/ excited 
states proceeds as follows. The ratio of weights 
in the Monte Carlo procedure is no longer a simple ratio of determinants
(eqn. (\ref{rat})). The ratio of weights for 
Ising-configurations $\{s^{\prime}\}$ and $\{s\}$ takes the form,
\begin{equation}
{{\omega^{\Gamma}(\{s^{\prime}\})} \over  {\omega^{\Gamma}(\{s\})} }
~~=~~    
%\sum\limits_{i=1}^{p} \sum\limits_{j=1}^{p} c_i c_j 
%\prod_{\sigma} 
{{  \langle L^{\Gamma}(l,\{s^{\prime}\})|
R^{\Gamma}(l,\{s^{\prime}\}) \rangle }
\over
 { \langle L^{\Gamma}(l,\{s\})|
R^{\Gamma}(l,\{s\}) \rangle }}
\label{omegaprime}
\end{equation} 
\noindent
where the projected states $ |R^{\Gamma}(l,\{s_{R}\}) \rangle$ 
and $\langle L^{\Gamma}(l,\{s_L\}) |$ are also  linear combinations of
states, 
\begin{equation}
|R(l,\{s_{R}\}) \rangle ~~=~~
\sum\limits_{j} c_{j}^{\Gamma}\prod\limits_{\sigma}
 |R_{\sigma}^{j\Gamma}(l,\{s_{R}\})\rangle
 \label{rketmulti}
\end{equation}
\begin{equation}
\langle L(l,\{s_{L}\})| ~~=~~
\sum\limits_{j} c_{j}^{\Gamma}\prod\limits_{\sigma}
\langle L_{\sigma}^{j\Gamma}(l,\{s_{L}\}) |
\label{lketmulti}
\end{equation}
\noindent
with each state in the summations  obtained in a manner
analogous to the single-determinantal case. The ratio
(eqn. (\ref{omegaprime})) is now given as the ratio of sums of determinants
appearing in the numerator and the denominator. 
\begin{equation}
{{\omega^{\Gamma}(\{s^{\prime}\})} \over  {\omega^{\Gamma}(\{s\})} }
=
{{ \sum\limits_{ij} \prod_{\sigma}c^{\Gamma}_{i} c^{\Gamma}_{j}
det \Bigl( {\bf{L}}_{\sigma}^{i\Gamma}(l,\{s^{\prime}\})
{\bf{R}}_{\sigma}^{j\Gamma}(l,\{s^{\prime}\}) \Bigr) }
\over
{ \sum\limits_{ij}\prod_{\sigma} c^{\Gamma}_{i} c^{\Gamma}_{j}
det \Bigl( {\bf{L}}_{\sigma}^{i\Gamma}(l,\{s\})
{\bf{R}}_{\sigma}^{j\Gamma}(l,\{s\}) \Bigr) }}
\label{omegadet}
\end{equation} 
\noindent
Evaluating the
ratio hence turns out to be more time consuming than in the single-
determinantal case. The expectation value of a single-particle operator 
$\hat{O}_{\sigma}$  can be obtained from an importance sampling
procedure, as in the single-configurational case,  with 
\begin{eqnarray}
O^{\Gamma}_{\sigma}(\{s\})=
{{\sum\limits_{i,j=1}^{p}  c_i^{\Gamma} c_j^{\Gamma}
%\prod_{\sigma} 
  \langle L^{j\Gamma}_{\sigma}(l,\{s\})| \hat{O}_{\sigma}|
R^{i\Gamma}_{\sigma}(l,\{s\}) \rangle 
\langle L^{j\Gamma}_{-\sigma}(l,\{s\})| 
R^{i\Gamma}_{-\sigma}(l,\{s\}) \rangle 
}
\over
{\sum\limits_{i,j=1}^{p}  c_i^{\Gamma} c_j^{\Gamma}
%\prod_{\sigma} 
  \langle L^{j\Gamma}_{\sigma}(l,\{s\})|
R^{i\Gamma}_{\sigma}(l,\{s\}) \rangle 
\langle L^{j\Gamma}_{-\sigma}(l,\{s\})| 
R^{i\Gamma}_{-\sigma}(l,\{s\}) \rangle 
}}
\label{Omulti}
\end{eqnarray}
\noindent
Estimates of two-particle properties can be obtained using
Wick's theorem. The expectation value of a two-particle
operator $\hat{Q}$ $=\hat{O}_{\sigma} \hat{O}_{-\sigma}$
can be expressed as
\begin{eqnarray}
Q^{\Gamma}(\{s\})=
{{\sum\limits_{i,j=1}^{p}  c_i^{\Gamma} c_j^{\Gamma}
%\prod_{\sigma} 
  \langle L^{j\Gamma}_{\sigma}(l,\{s\})| \hat{O}_{\sigma}|
R^{i\Gamma}_{\sigma}(l,\{s\}) \rangle 
\langle L^{j\Gamma}_{-\sigma}(l,\{s\})| \hat{O}_{-\sigma}|
R^{i\Gamma}_{-\sigma}(l,\{s\}) \rangle 
}
\over
{\sum\limits_{i,j=1}^{p}  c_i^{\Gamma} c_j^{\Gamma}
%\prod_{\sigma} 
  \langle L^{j\Gamma}_{\sigma}(l,\{s\})|
R^{i\Gamma}_{\sigma}(l,\{s\}) \rangle 
\langle L^{j\Gamma}_{-\sigma}(l,\{s\})| 
R^{i\Gamma}_{-\sigma}(l,\{s\}) \rangle 
}}
\label{twopartmlt}
\end{eqnarray}
Property estimates in the MSPQMC procedure can be carried out as in
the single configurational PQMC procedure. Estimates have been
obtained at an extreme time-slice (usually chosen to be the last), 
at the middle-time slice and by averaging over all time-slices.
We use the orthogonalization procedure on each of the left and 
right projected configurations in the multi-configurational
trial wavefunction, every few time steps.  Likewise, the Green
function updating algorithm can be applied independently to  each term
in eqn. (\ref{Omulti}).
We have carried out single- and multi-configurational PQMC simulations
using the three averaging procedures for the Green function and
other properties described above. An analysis of the errors
involved in these procedures is presented in the next subsection.

\subsection{Validity of the MSPQMC method and error analysis}

While we have described a method for carrying out Monte Carlo
simulations of excited states, the validity of such a procedure
needs to be examined.
In the MSPQMC method the trial wavefunction is a symmetrized linear
combination of electron configurations.
If the projection  ansatz is employed 
without any further approximations, beyond employing a finite
$\beta$, the projected state would
continue to be in the initially chosen symmetry space. Employing a
Trotter decomposition followed by a H-S transformation yields
the approximation for the density operator $exp(-\beta \hat{H})$
given in eqns. (14,15). This procedure still 
retains the symmetry
property of the density operator. However, in the Monte Carlo 
procedure, we only sample a fraction of all possible Ising-
configurations. The operator $\hat{W}(\{s\})$ for an arbitrary
Ising-configuration $\{s\}$ does not have the symmetry of the
Hamiltonian. 

Thus, it appears that the estimated properties would
also have contributions from states belonging to symmetry subspaces other
than the chosen space. In such a case, it is not enough to choose
the symmetry of the trial wavefunction but is also necessary to
prevent admixture with states of other symmetry in the MC procedure.
While targetting the ground state, such an admixture would
only slow down the convergence to the ground state since the 
admixture could involve intruder states of other symmetries
lying between the ground 
state and the first excited state of the symmetry subspace of
the ground state. On the other hand, while targetting excited states,
the admixture could lead to intrusion of the ground state and projection 
would eventually lead to the ground state.

However, it is possible to retain the symmetry of the 
Hamiltonian, even when all the Ising-configurations are 
not sampled, using the following procedure. The set of Ising-configurations
can be divided into disjoint invariant subsets. Any Ising-
configuration in an invariant subset can be generated from any
other configuration in the same subset by operating with an 
element $\hat{R}$ of the Schr\"odinger group. The operator 
$\hat{W}^{sym.}$,
defined as
\begin{eqnarray}
\hat{W}^{sym.}(\{s\}) = \sum\limits_{\hat{R}}
\hat{W}(\hat{R}\{s\}),
\label{wsym}
\end{eqnarray}
\noindent
has the symmetry of the Hamiltonian. The estimates obtained
from the projection procedure carried out using this symmetrized 
operator $\hat{W}^{sym.}$ exclude contributions from symmetries 
other than that of the initially chosen subspace.
In the estimation of energy, the explicit use of the symmetrized
operator, $\hat{W}^{sym.}$, is unnecessary for the following reason. 
The Hamiltonian of the system can be expressed as a linear combination
of terms obtained by operating with
the symmetry operators of the group on an irreducible operator,
$\hat{H}^{irr.}$, which contains terms such as nearest-neighbour 
transfer operators and site-diagonal interactions to yield,
\begin{eqnarray}
\hat{H} = \sum\limits_{\hat{R}} 
\hat{R} \hat{H}^{irr.}
\label{hunq}
\end{eqnarray}
\noindent
For example, for a Hubbard ring of $N$ sites with periodic
boundary conditions, $\hat{H}^{irr.}$ is
given by
\begin{eqnarray}
\hat{H}^{irr.} = 
( t_{i,i+1} \hat{a}^{\dagger}_{i\sigma} \hat{a}_{i+1\sigma} + h.c.)
+ U\hat{n}_{i \uparrow}\hat{n}_{i \downarrow},
\label{unqN}
\end{eqnarray}
\noindent
where $i$ is an arbitrary site.
The Hamiltonian can be generated from $\hat{H}^{irr.}$
by adding up terms obtained by successive ${{2\pi}\over{N}}$ rotations.
For a system with a more complicated topology and non-equivalent
nearest-neighbour bonds, such as $C_{60}$ with bond-alternation,
$\hat{H}^{irr.}$ is given by
\begin{eqnarray}
\hat{H}^{irr.} = 
-{{1}\over{2}} t_{hh} (\hat{a}^{\dagger}_{i\sigma} \hat{a}_{j\sigma} + h.c.) -
 t_{hp} (\hat{a}^{\dagger}_{i\sigma} \hat{a}_{k\sigma} + h.c.)
+{{1}\over{2}} U\hat{n}_{i \uparrow}\hat{n}_{i \downarrow}.
\label{unqc60}
\end{eqnarray}
\noindent
where $t_{hh}$ is the transfer integral for a hexagon-hexagon bond and
$t_{hp}$ is the transfer integral for a hexagon-pentagon bond. The
label $i$ corresponds to an arbitrary site on the truncated icosahedral
lattice and $j$ and $k$ label nearest-neighbour sites corresponding
to hexagon-hexagon and hexagon-pentagon bonds.
The various terms in the Hamiltonian can be generated from $\hat{H}^{irr.}$
by operation with all the 120 elements of the icosahedral group.
The estimate of the Hamiltonian for a single Ising-configuration
in the MSPQMC procedure is given by
\begin{eqnarray}
E^{\Gamma}(l,\{s\}) = \sum\limits_{\hat{R}} 
 \langle \phi^{\Gamma} | \hat{W}(\{s_{L}\})
\hat{R} \hat{H}^{irr.}
\hat{W}(\{s_{R}\}) |\phi^{\Gamma} \rangle,
\label{emspqmc}
\end{eqnarray}
\noindent
which can be rearranged to yield,
\begin{eqnarray}
E^{\Gamma}(l,\{s\}) = \sum\limits_{\hat{R}} 
 \langle \phi^{\Gamma} | {\hat{H}}^{irr.}
\sum\limits_{\hat{R}} {\hat{R}}
{\hat{W}}(\{s\}) |\phi^{\Gamma} \rangle,
\label{emspqmc1}
\end{eqnarray}
\noindent
in doing which we have incurred a Trotter error. However, the operator
acting on the ket conserves the symmetry of the initial state. The 
evaluation of energy on the last time slice is
{\em{a priori}} symmetrized and does not involve the additional Trotter
error.

For operators that do not have the symmetry of the Hamiltonian, we need to
explicitly enforce the symmetrization of the estimates, even at the last 
time slice. By this, we would ensure that any property estimates would
correspond to the excited state, targetted as the lowest state in a 
given symmetry subspace. This is done using the following procedure\cite
{pqmcc60}.
When a particular Ising-configuration is
visited in the course of sampling, property estimates are obtained for
all Ising-configurations related by symmetry.
The symmetry of the Hamiltonian guarantees that the one--step
transition probability between Ising-configurations $\{s\}$,
$\{s^{\prime}\}$ is the same as that between  $\hat{R}\{s\}$ and
$\hat{R}\{s^{\prime}\}$, 
\begin{equation}
p_{\{s\}\rightarrow\{s^{\prime}\}} =  p_{\hat{R}\{s\}\rightarrow\hat{R}
\{s^{\prime}\}}
\end{equation}
\noindent
Therefore, if an Ising configuration $\{s^{\prime}\}$ is accepted (rejected) 
from an
initial configuration $\{s\}$, the same result is expected from all symmetry
related pairs of configurations $\hat{R}\{s^{\prime}\}$ and $\hat{R}\{s\}$.
This feature can be incorporated by  constructing a symmetrized Green
function as follows:
\begin{equation}
({\bf{G}}^{sym}_{\sigma}(l,\{s\}))_{mn} = {1\over h} \sum\limits_{ {\hat{R}}}
({\bf{G}}_{\sigma}({\hat{R}}\{l,s\}))_{mn}
\end{equation}
\noindent
and $\hat{R}$ runs over all the $h$ symmetry elements of the group.

It appears from the equation that we need to update the Green functions
${\bf{G}}_{\sigma}(\hat{R}\{s\})$ for every symmetry operation $\hat{R}$, which could 
be enormously computationally intensive. However, the Green functions
${\bf{G}_{\sigma}}(\hat{R}\{s\})$ and ${\bf{G}}_{\sigma}(\{s\})$ are related as
\begin{equation}
({\bf{G}}_{\sigma}(l,\{s^{\prime}\}))_{mn} = 
({\bf{ G}}_{\sigma}(l,\{s\}))_{m^{\prime}n^{\prime}}
\end{equation}
\begin{equation}
\hat{R} :   \{s\}  \rightarrow \{s^{\prime}\};
\end{equation}
\begin{equation}
   \hat{R}^{-1} :  i \rightarrow  i^{\prime} \hspace{0.2cm}  ; 
\hspace{0.2cm}   \hat{R}^{-1} :  j \rightarrow  j^{\prime}
\end{equation}
\noindent
Thus, from the Green function of a  single Ising configuration, we can generate
the Green function of all Ising configurations related by the Schr\"odinger 
group of the system and thus ensure that property estimates are obtained
in the irreducible representation $\Gamma$ even for excited states.

The discussion so far seems to indicate that it is accurate to 
estimate the energy and other properties at an extreme time-slice
since we do not incur the additional Trotter error.
To analyze the errors arising from the updating and averaging
procedures, we recognize that
the ratio $r$ (eqn. (\ref{rat})) is independent of the time-slice
$l$ at which the lattice is notionally partitioned into
left and right halves in both the single- and multi-
configurational procedures. However, the matrix elements of the
Green function depend on the time-slice at which they are
computed.  
As seen from eqn. (\ref{psiapprox}), {\em both} the
right and left projected states are good approximations
to the ground state only when they have both been sufficiently 
evolved, i.e. when $l \approx L/2$.  Thus, we expect that the
time-slice at which the averaging is carried out also determines 
the accuracy of the estimated properties. Because, if either
the trial state or its transpose is insufficiently projected,
contributions due to admixture with excited states of the same
symmetry would
lead to inaccurate estimates. For operators $\hat{O}$ that 
commute with the Hamiltonian, 
the ground-state expectation  value can be obtained accurately 
{\em {even}} at the last time slice
(for a non-degenerate ground-state) since,
\begin{eqnarray}
 {{\langle \phi| \hat{O} |\psi_{0}\rangle }
\over
{ \langle \phi | \psi_{0}\rangle}} =
O_{00} 
\label{O00}
\end{eqnarray}
\noindent
where
\begin{eqnarray}
O_{ij} = \langle \psi_{j} | \hat{O} | \psi_{i} \rangle \\
|\phi\rangle = \sum\limits_{k} a_k |\psi_k\rangle
\end{eqnarray}
\noindent
and $|\psi_k\rangle$ are the eigenstates of the Hamiltonian. 
However, when $\hat{O}$ does not commute with the Hamiltonian,  an estimate
of its expectation value,  carried out at
the last time-slice would yield,
\begin{eqnarray}
 {{\langle \phi| \hat{O} |\psi_{0}\rangle }
\over
{ \langle \phi | \psi_{0}\rangle}} =
O_{00} + {1\over a_{0} } \sum\limits_{k \ne 0} a_k O_{k0}
\label{Ok0}
\end{eqnarray}
\noindent
Thus the estimates of such properties are prone to be rather
inaccurate towards either end of the Ising lattice.  From this analysis
we expect that the energy is most accurately estimated at an extreme
time-slice, while other properties that do not have the symmetry of the
Hamiltonian should be estimated at $l \approx L/2$.
We have compared estimates obtained as time-slice averages 
and those obtained from measurements at a single time-slice
with exact results, which we present in the next section.

In the MSPQMC method for excited states, we encounter the
negative sign problem even at half-filling although the number of
occurrences of negative signs even at large $U/t$ is insignificant.
In the usual quantum Monte Carlo methods, the sign problem arises
only away from half-filling for
bipartite lattices in any dimension. For, at half-filling in
a bipartite lattice, the determinants
for the up and down spins can be shown to have the same sign, if they
occupy the same set of molecular orbitals. 
In the MSPQMC method, the sign problem could arise for two
additional reasons. The individual up and down spin determinants
could have different signs if the MOs occupied by the up and down 
spin electrons are not identical.
Besides, the phases with which the configurations in the trial state are
combined could also produce an overall negative sign even if the
products of individual determinants of 
up and down spin corresponding to 
$\langle L_{\sigma}^{j\Gamma}(l,\{s_{L}\}) | 
R_{\sigma}^{j^{\prime}\Gamma}(l,\{s_{R}\}) \rangle$ are positive. 
In what follows, we present data to show that the additional negative
signs arising from the different up- and down-spin configurations
as well as the phases are a negligible fraction. Besides, even the
absolute numbers of negative signs appears to decrease with increasing
system size for the excited states of half-filled systems.

\section{Results and Discussion}

In this section we first present our MSPQMC results on the ground and excited
state properties of one-dimensional Hubbard systems at half-filling. We
compare these with results from exact diagonalization studies, both for
energies and correlation functions. We also report results of
studies carried out on the ground state of doped Hubbard systems and
compare the binding energy obtained from these studies with Bethe 
ansatz results. We then report results on the excited states of the
Hubbard chains. Results for the two-dimensional lattice include the
$4 \times 2$ ladder and the $3 \times 4$  lattice. The 
$4 \times 2$ and $3 \times 4$ lattices are amenable to
exact diagonalization studies and hence we have extensively studied
various properties of these lattices for states of different symmetries
at hole doping ranging from $2$ to $4$ holes. 

In all these studies, we focus
on two sources of error common to all PQMC procedures.
The first one is that the projection as implemented via
H-S fields does not retain the symmetry of the initial state for
individual Ising configurations. However, it can be shown 
that the symmetry is retained if estimates are carried
out at the last time slice. Except for the estimation of energy,
estimates carried out in this manner contain contributions from excited
states of the same symmetry. The error due to this could perhaps be 
reduced to some extent by the choice of trial wavefunction.  
As discussed before, there is an additional error of the order 
of Trotter error that is incurred by resorting to property estimates 
at intermediate time slices or by averaging over all
time slices, as in the single configurational PQMC algorithms. In
these procedures, however, the excited states are better filtered out due to 
projection being carried out on the trial state as well as its transpose.
We have systematically studied the properties
of the system obtained from (i) estimation at the last time slice (L)
(ii) by averaging over estimates at all time
slices (A) and (iii) estimation at the middle time slice (M).
 We have compared the results for energies and other
correlation functions with exact results. We find that the energy estimates
are most accurate on the last time slice while the estimates of 
correlation functions are accurate at the middle time slice. However,
we find that the correlation functions at the last time slice for
the chosen trial  wavefunctions are reasonably accurate. More
importantly, the accuracy of the estimates obtained at the middle time-slice
and by averaging
over all time slices can be improved by tuning the projection parameter
which is consistent with the view that the errors in these two schemes
are Trotter-like.

\subsection{Ground state properties at half-filling for 1-D systems}

We have computed the ground state properties of Hubbard rings at 
half-filling 
for various values of $U/t$, using
different PQMC averaging procedures to find out the best suited
algorithms for energies and other properties. The projection
parameter $\beta$ was set at 2.0 and $\Delta\tau$ was fixed at $0.05$ for
$U/t \geq 4.0$ and at $0.1$ for smaller values of $U/t$.

In Table 1, we present the enrgies of rings of six and fourteen
sites for various values of $U/t$. The agreement between exact and
PQMC energies is very good for small and intermediate values 
of $U/t$ from all the three averaging procedures. We note, howver, that
for the larger ring averaging at the last time-slice
gives best energy estimates. In Table 2, we compare, for the ring of
six sites, spin correlations obtained from the three averaging
procedures in PQMC with those obtained from exact, variational 
Monte Carlo and symmetrized PQMC (with Green function updating)
calculations. The PQMC estimates from middle-time slice as well
as from averages over all time slice agree remarkably well with
exact results. Green function updating, with judicious recomputation,
yields results that are very similar to those obtained from the
computationally more expensive explicit recomputation scheme 
followed in the all-time slice averaging procedure. In Table 3 
we present the spin-spin and charge-charge correlation
functions for weak, intermediate and strong electron-correlations.
The charge and spin correlation even for strong electron correlations 
are well reproduced. Here again, we note that the average and middle
time-slice estimates are better than the last time-slice estimates.

\subsection{Excited state properties at half-filling for 1-D systems}

In this subsection, we study the behaviour of the
"optical" and "spin" gaps of Hubbard chains with increasing
strength of electron correlations and compare these with exact results. 
The energies presented are
obtained from the different averaging schemes. In an earlier
paper, we reported some of the excitation gaps which were, however,
obtained from energies calculated only at the last time-slice.
The emphasis of this subsection is on comparing the various
quantities computed using the different averaging schemes described earlier
in this paper.
Besides energies, these comparisons include various excited state
correlation functions
which have been studied for the first time using the MSPQMC procedure.

In Table (4), we present MSPQMC energies for the singlet excited states
of Hubbard chains at weak and intermediate correlation strengths. We
use a larger projection parameter in schemes (A) and (M). We
observe that at small system sizes, the energies obtained from procedures
(A) and (M) have slightly larger errors than those obtained from averaging
at the last time-slice. Furthermore, the differences in estimates obtained 
from schemes (L), (A) and (M) decrease with increasing system size. 
We find that the estimates of the triplet excited state energy of the 
chain of 14  obtained from all the procedures ((L), (A) and (M)) are comparable. 
In Table (5) we present the
diagonal and longer-range spin correlations in the singlet excited
state of the chain of 6, for $U/t=2.0$ and $6.0$.  Here, we find that the
average time slice (A) and the middle time slice (M) values have smaller 
errors compared to the values calculated at the last time slice (L).
This trend also holds for the charge correlations.

The above results show that it is indeed possible to obtain good 
estimates of {\em excited 
state} energies and other properties by averaging 
over all time-slices, if the Trotter error is controlled by sufficiently 
fine time-slicing. The wide applicability of the MSPQMC method would 
depend critically on the viability of this averaging method.
For, it allows the use of the $O(N^2)$ Green function updating algorithm
and thus makes larger system sizes computationally accessible. 

\subsection{Properties of doped 1-D systems}

In this subsection we study Hubbard rings away from
half-filling using the MSPQMC method. The quantity of interest in these
systems is the pair binding  energy of holes. The results presented here 
have been obtained using the Green function updating algorithm.
Fye, Martins and Scalettar\cite{fyehbe} obtained the pair binding
energies for holes in doped Hubbard rings using the Sutherland-Shastry 
generalization of the Bethe ansatz equations for arbitrary boundary 
conditions.  The binding energy for two holes in a 
system of $N$ electrons is defined as $E(N)+E(N-2)-2E(N-1)$.
They found that the binding-energies show non-monotonic behaviour with system
size. The computation of these pair binding energies provides a
very stringent test of any numerical scheme, as
these quantities are small differences of relatively large numbers.

In Fig. (1) we plot the pair binding energies of periodic $4n$ and $4n+2$ 
Hubbard rings against correlation strength. Our data compare 
well with the Bethe ansatz data, reproducing the important qualitative features.
As observed earlier\cite{fyehbe}, $4n+2$ systems doped with two holes do not 
show binding while $4n$ systems do exhibit negative binding energies 
over a certain parameter regime, in agreement with the Bethe ansatz results.

We now turn our attention to the negative sign problem for 
"open shell"/excited states of one-dimensional
systems which arises due to the reasons discussed earlier.
In the triplet state, the negative sign could arise either from
non-identical occupancies of MOs
for up and down spin electrons in the trial wavefunction or 
from the phase with which the terms in the 
trial wavefunction are combined.
As seen in Table (6) the number of negative signs even 
for the excited triplet is not large. In fact, the 
number of occurances is a neglible fraction of the total number
of configurations sampled and this decreases with increasing 
system size for the systems studied.

\subsection{Properties of the two-dimensional Hubbard model}

While it is possible to obtain properties of the Hubbard model
on one-dimensional lattices from a variety of methods, both analytical
and numerical, the two-dimensional Hubbard model has proved much
harder to study. In this subsection, we illustrate the power of
the MSPQMC method by studying in detail the $4 \times 2$  and the 
$3 \times 4$ clusters. These systems are easily amenable to 
exact diagonalization studies and provide the necessary checks.
In Table (7) we present the energies of the $4 \times 2$ ladder with 
6 electrons and the $ 3 \times 4$ system with 8 electrons.
These fillings have open-shell non-interacting ground states. 
The $4 \times 2$ lattice has a triply degenerate HOMO while
the $3 \times 4$ lattice has a doubly degenerate HOMO 
at the chosen fillings. The MSPQMC method accurately resolves the 
singlet and the triplet of the $4 \times 2$ and the $3 \times 4$ 
lattices with a trial wavefunction which is a symmetrized linear
combination of properly chosen Slater determinants. We have also computed
the ground state of the half-filled Hubbard Hamiltonian on the $4 \times 4$
lattice. In the non-interacting limit, this system has a six-fold 
degenerate MO at the Fermi level. However, a muti-configurational 
trial wavefunction which is an appropriate symmetrized linear 
combination of just two MO-configurations yields a ground state energy
which differs from the exact result by $1.4 \%  $ for $U/t = 4.0$.

In Table (8), we study the effect of the three averaging schemes, (L),
(A) and (M) described previously on the energy of  the $3 \times 4$
lattice with 8 electrons at various values of the correlation strength.
We used larger projection parameters for schemes A and M.  As expected
from our earlier analysis, we once again find that energies obtained at 
the last time slice are more
accurate, but suitable tuning of the projection parameter allows us to
reproduce energies from schemes A and M with comparable accuracy. This
feature of the averaging schemes A and M is important,
since we expect these two schemes to provide 
accurate correlation functions. In Fig. (2), we present the hole-binding
energies for four holes on the $3 \times 4$ lattice. It is interesting
to note that the $3 \times 4 $ lattice does not show binding in the
parameter regime studied. We also present the energy difference between
the high and low spin states of 8 electrons on the $3 \times 4$ cluster
and note that Hund's rule is obeyed over at weak and also in the 
intermediate correlation regime.  Apart from the energy, other correlation 
functions like the spin-spin and charge-charge correlations charecterize 
the state of the system.  We prefer to study these correlation functions
for the $4 \times 2$ ladder, with bond-alternation, where the transfer
integral of the rung, $t_{rung} = 0.9 t$, where $t$ is the transfer integral
between two nearest-neighbours on each leg of the ladder. This choice of
system and tranfer integrals has been made to ensure that the state studied
is non-degenerate. In Table (9) we present the singlet spin correlations of 
the bond-alternated $4 \times 2$ ladder with 8 electrons obtained from
the three averaging schemes, (L), (A) and (M). We observe that as expected
from our analysis, correlation functions obtained from averaging at the
middle time slice are significantly better than those obtained at the
last time slice, both for weak and for intermediate correlation
strengths. However,
correlations obtained by averaging over all time slices, using the
Green function updating algorithm are seen to be almost as accurate as
those obtained from procedure (M). Thus, we expect that this method can be
used to study much larger lattices. The charge correlations, also presented
in Table (9) are seen to follow similar trends.

A major hindrance in the application of quantum Monte Carlo methods to 
the Hubbard model at large correlation strength and/or away from half-filling
is the so-called 'negative-sign' problem. We have discussed the additional
sign problem that arises even for one-dimensional systems when excited
states are targetted. Bipartite lattices away from half-filling in 2-D suffer
from the negative sign problem even when the ground state is the
targetted state. Lattices which do not have charge-conjugation symmetry
can have negative ratios of determinants even at half-filling. As discussed
in the one-dimensional case, a multi-configurational trial wavefunction 
can lead to an additional sign problem when the configurations of electrons
of up and down spin are not identical as well as through the phases with
which terms in the wavefunction are combined. In Table (10), we present
the actual numbers of negative signs we encountered in simulating the
$4 \times 2$ and the $3 \times 4$ lattices with 6 and 8 electrons respectively.
We observe that we are able to reduce the number of occurances of
negative signs by a suitable
choice of projection parameter and that a muti-configurational trial 
wavefunction does not appear to significantly worsen the sign problem in
simulations of the two-dimensional Hubbard model, even of non-bipartite
lattices.

\section{Summary}

We have described in detail the procedure for obtaining ground and
excited states of open-shell systems, using a trial muti-configurational
wavefunction within the projector quantum Monte Carlo method. A careful
analysis of the method for excited states
leads naturally to the idea of symmetrized sampling for 
correlation functions, developed earlier in the context of ground state
simulations. It also leads to three possible averaging schemes, in which
property estimates are carried out 
at the last time slice, over all time slices and at the middle time slice.
We have analyzed the errors incurred in these various averaging procedures.
From these analyses, we expect that the energy is best estimated at the
last time slice.  We also expect that the error incurred in the other
procedures is Trotter like and can be reduced by increasing the projection
parameter. Correlations that do not have the full symmetry of the
Hamiltonian are better estimated at the middle time slice. We find 
that the energies and spin and charge correlations of
one- and two-dimensional lattices, at and away from half-filling
do exhibit this behaviour. We also find that upon increasing the projection
parameter, properties obtained by averaging over all time slices and by 
averaging at the middle time slice have comparable accuracy. This
observation allows
the use of a Green function updating algorithm and makes larger 
system sizes accessible by the MSPQMC method. 
We have used this
technique to study the hole-binding energies oftwo holes in $4n$ and $4n+2$ 
systems, which compare well the Bethe ansatz data of Fye, Martins and
Scalettar. We have also studied small clusters
amenable to exact diagonalization studies in 2-D and have reproduced 
their energies and correlation functions by the MSPQMC method.
We identify two ways in which a 
multiconfigurational trial wavefunction can lead to a negative
sign problem. We observe that this effect is not severe in 1-D and tends to
vanish with increasing system size. We also note that this does not 
enhance the severity of the sign problem in two dimensions. The MSPQMC
method has been demonstrated to be capable of yielding
reliable properties of ground and low-lying excited states
of the Hubbard model.

\noindent
{\bf{Acknowledgement:}} We wish to thank B. S. Shastry for useful discussions.
We thank B. S. Shastry and M. J. Martins for the Bethe ansatz
data used for comparison. B. S. 
wishes to thank R. M. Fye and M. J. Martins for useful discussions.
We thank  Biswadeb Dutta for help with the computer systems at the JNCASR 
and Y. Anusooya for generating some of the exact results. 
We thank the SERC, IISc for time on the 8-node DEC Turbo Laser
and the JNCASR for the use of the 4-node R10000 Power Challenge.

\pagebreak

\pagebreak
\clearpage

\begin{table}
\begin{center}
\caption{MSPQMC energies of rings of 6 and 14 sites compared with 
exact calculations for averaging at last (lts), middle (mts) and all (ats)
time slices for $U/t = 2.0, 4.0$ and 6.0. 
Data for $U/t = 4.0$ and $6.0$ is with $\Delta\tau = 0.05$. Numbers in
parantheses are the orders of magnitude of the statistical error. Data from
procedures (A) and (M) have similar errors.}
\begin{tabular}{|c|cccc|cccc|}   \hline
\multicolumn{1}{|c|}{$U/t$}& \multicolumn{4}{c|}
{Energy-ring of 6}  &\multicolumn{4}{c|} {Energy-ring of 14}\\ \hline
&   Exact  & lts &ats& mts & $Exact$  & lts &ats& mts \\ \hline
1.0 & -6.601 &-6.598& -6.605& -6.609& -14.715& -14.710($10^{-4}$) & -14.687& 
 -14.718\\
4.0 & -3.669 &-3.657& -3.703& -3.685 & -8.088 & -7.976($10^{-4}$)& -8.247&  -8.246\\
6.0 & -2.649 &-2.599&  -2.704&   -2.661& -5.916 & -5.875($10^{-3}$) & -6.018 & -5.983\\
10.0& -1.664 &-1.525& -1.785& -1.768& -3.763 &  -3.674($10^{-2}$)&  
 -4.004 & -4.077\\ \hline
\end{tabular}
\end{center}
\end{table}

\begin{table}
\begin{center}
\caption{ Spin--Spin correlations ($4\langle s_{i}^{z}s_{j}^{z}\rangle$) 
of benzene for $U/t = 1.0$ and $4.0$, from
exact and MC calculations. The MC data is from variational
MC (VMC), symmetrized PQMC (SPQMC) with green function updating and
MSPQMC with explicit recalculation of the green function, followed by
averaging at the last time slice (lts),
over all time slices (ats) and at the middle time slice
(mts). Data for $U/t = 4.0$ is with $\Delta\tau = 0.05$.}
\begin{tabular}{|c|c|cccccc|}   \hline
\multicolumn{1}{|c|}{$U/t$}&\multicolumn{1}{c|}{$i,j$} & \multicolumn{6}{c|}
{$4\langle s_{i}^{z}s_{j}^{z}\rangle$}  \\ \hline
& &   Exact & VMC& SPQMC  & lts &ats& mts \\ \hline
1.0  &1,1 &  0.567 &  0.557   & 0.562   &  0.533 &0.562
& 0.568\\
1.0  &1,2 & -0.267 & -0.245 & -0.264   &-0.243  &-0.262 
& -0.266\\
1.0  &1,3 &  0.022 & 0.011   & 0.020    & 0.009  &0.018
& 0.021\\
1.0  &1,4 & -0.077 & -0.062 & -0.074   &-0.064  &-0.073
&-0.076 \\ \hline
4.0  &1,1 & 0.778  & 0.750  & 0.751     & 0.638  &0.748
& 0.774\\
4.0  &1,2 &-0.435  & -0.395  & -0.412   &-0.323  &-0.412 
&-0.433\\
4.0  &1,3 & 0.140  & 0.077 & 0.120     & 0.060  &0.122
& 0.140\\
4.0  &1,4 &-0.188  & -0.113  & -0.167   &-0.113  &-0.169
 &-0.186\\\hline
\end{tabular}
\end{center}
\end{table}

\begin{table}
\begin{center}
\caption{ Spin-Spin correlations ($4\langle s_{i}^{z}s_{j}^{z}\rangle$) 
and charge-charge correlations ($4\langle n_{i}^{z}n_{j}^{z}\rangle$) 
of the 6-site Hubbard ring
exact and MC calculations. The MSPQMC values have been
obtained by averaging at the last time slice (lts),
over all time slices (ats) and at the middle time slice
(mts). Data for $U/t = 4.0$ and $6.0$ is with $\Delta\tau = 0.05$.}
\begin{tabular}{|c|c|cccc|cccc|}   \hline
\multicolumn{1}{|c|}{$U/t$}& \multicolumn{1}{c|}{$i,j$}&
\multicolumn{4}{c|}{Spin} &
\multicolumn{4}{c|}{Charge} \\ \hline
& & Exact & L &    A &    M & Exact  & L   & A   & M   \\  \hline
2.0 & 1,1 &  0.638 &  0.567 &  0.626 & 0.639
 &   1.362 &	  1.433 &  1.374 &1.361 \\
&1,2 & -0.320 & -0.267 & -0.308 &-0.317
    &  0.849 &	  0.814 &  0.844 &  0.850 \\
    & 1,3 &  0.055 & 0.023 &  0.042 & 0.047
    &  0.985 &  0.990 &  0.984 &  0.984 \\
    & 1,4 & -0.109 & -0.079 & -0.095 &-0.100 
    &  0.972 &  0.960 &  0.967 &  0.970 \\ \hline
6.0 & 1,1 &  0.873 &  0.700 & 0.844 &   0.873 
   &   1.127 &  1.300 &  1.159 &  1.127 \\
    & 1,2 & -0.516 & -0.378 &-0.496 & -0.520 
   &   0.942 &  0.868 &  0.935 &  0.943 \\
    & 1,3 &  0.202 & 0.103  &0.186  & 0.202 
     &   0.996 & 0.991 &  0.996 &  0.995 \\
    & 1,4 & -0.244 & -0.151 &-0.224 & -0.237 
     &   0.997 &  0.983 &  0.996 &  0.997 \\ \hline
\end{tabular}
\end{center}
\end{table}

\begin{table}
\begin{center}
\caption{MSPQMC energies of the excited singlet states of chains of 6  and
14 sites, compared with 
exact calculations for averaging at last (L), all (A) and middle (M)
time slices for $U/t = 2.0$ and $6.0$. 
Data for $U/t = 6.0$ is with $\Delta\tau = 0.05$.}
\begin{tabular}{|c|cccc|cccc|}   \hline
\multicolumn{1}{|c|}{$N$}& \multicolumn{4}{c|}{$U/t=2.0$} &
\multicolumn{4}{c|}{$U/t=6.0$} \\ \hline
&   Exact  & L & A & M & Exact  &  L & A &  M \\ \hline
6 & 3.0175 & -3.0130 & -3.0216 & -3.0192 & 1.9212 & 1.9283 & 1.8874 & 1.9133 \\
8 &4.9958&-4.9893 &-5.0010 &-5.0044 &0.6990 & 0.7405 & 0.6853 & 0.6259\\
10 &  -6.8718&-6.8615 & -6.8645 & -6.8502& -0.3723 & -0.3477 &-0.4378&-0.3794 \\
12& 8.6916 & -8.6825 & -8.6994 &-8.7125 & -1.3639& -1.3415&-1.4977 & -1.4133 \\
14 &-10.4774 &-10.4605 &-10.4727 & -10.4651& -2.3089 &-2.2349 &-2.4093&
-2.4794 \\ \hline
\end{tabular}
\end{center}
\end{table}

\pagebreak
\clearpage

\begin{table}
\begin{center}
\caption{MSPQMC spin and charge correlations of the excited singlet states 
of the chain of 6, compared with 
exact calculations for averaging at last (L), all (A) and middle (M)
time slices for $U/t = 2.0$ and $6.0$. 
Data for $U/t = 6.0$ is with $\Delta\tau = 0.05$.}
\begin{tabular}{|c|c|cccc|cccc|}   \hline
\multicolumn{1}{|c|}{$U/t$}&\multicolumn{1}{c|}{$i,j$}&
\multicolumn{4}{c|}{spin} &
\multicolumn{4}{c|}{charge} \\ \hline
& & Exact & L &    A &    M & Exact  & L   & A   & M   \\  \hline
2.0& 1,1&  0.498& 0.431&  0.477 &   0.490 & 1.502 & 1.569 & 1.535 & 1.510 \\
& 1,2& -0.248&-0.231& -0.251 &  -0.256 & 0.861 & 0.805 & 0.857 & 0.855 \\
& 1,3& -0.157&-0.124& -0.136 &  -0.152 & 1.017 & 1.052 & 1.048 & 1.021 \\
& 1,4&  0.020& 0.005&  0.013 &   0.018 & 0.871 & 0.845 & 0.868 & 0.868 \\
& 1,5& -0.043&-0.034& -0.040 &  -0.043 & 0.939 & 0.978 & 0.957 & 0.947 \\
& 1,6& -0.071&-0.046& -0.056 &  -0.057 & 0.810 & 0.750 & 0.792 & 0.799 \\
 \hline
6.0 & 1,1&  0.644 & 0.513 &  0.596 & 0.603   & 1.356 & 1.487 & 1.409 & 1.397 \\
& 1,2& -0.292 & -0.266 & -0.292 & -0.295 & 0.952 & 0.878 & 0.943 & 0.943 \\
& 1,3& -0.242 & -0.172 & -0.214 & -0.219 & 0.980 & 1.015 & 0.993 & 0.977 \\
& 1,4&  0.056 &  0.027 &  0.026 & 0.025  & 0.899 & 0.862 & 0.885 & 0.883 \\
& 1,5& -0.038 & -0.033 & -0.024 & -0.027 & 0.906 & 0.935 & 0.913 & 0.910 \\
& 1,6& -0.129 & -0.070 & -0.088 & -0.087 & 0.907 & 0.823 & 0.884 & 0.890 \\
\hline
\end{tabular}
\end{center}
\end{table}
\pagebreak	
\clearpage

\begin{table}
\begin{center}
\caption{ Number of occurances of negative signs for "open-shell"/
excited states in one-dimension. Sample size is 
$N \times20\times5000$ for $U/t < 4.0$ and 
$N \times 40 \times 8000$ for $U/t \geq 4.0$.}
\begin{tabular}{|c|rc|rc|rc|rc|rc|rc|rc|rc|}   \hline
\multicolumn{1}{|c|}{$U/t$}&\multicolumn{2} {c|}{$N=6$}
 &\multicolumn{2}{c|} { $N=14$}  &\multicolumn{2}{c|} { $N=18$} 
 &\multicolumn{2}{c|} { $N=42$} \\ \hline
& singlet & triplet &singlet & triplet & singlet & triplet & singlet & triplet 
\\ \hline
 1.0 &    14   &  0  &0   &  0  & 0 & 0   &  0   &  0 \\
 2.0 &   532  &  14  &1   &  0  & 0 & 0   &  0   &  0 \\
 3.0 &   1568  &  184  &39  & 67 &21 & 15     &  0   &  0 \\
 4.0 &   2414  &  693  &224 & 371 & 71 & 192     &  0   &  0 \\
 5.0 &   3114  &  1131  &901 & 1328 & 397 & 857     &  0   &  17 \\
 6.0 &   3638  &  1416  &1325 & 1647  & 706 & 1085   &  0   &  127 \\ \hline
\end{tabular}
\end{center}
\end{table}

\begin{table}
\begin{center}
\caption{ MSPQMC singlet and triplet energies of the $4 \times 2$ ladder,
with 6 electrons, the $3 \times 4$ lattice with 8 electrons,
compared with exact calculations.}
\begin{tabular}{|c|c|c|cc|cc|}   \hline
\multicolumn{1}{|c|}{$U/t$}&\multicolumn{1}{c|}{system}
&\multicolumn{1}{c|}{$N_e$}&
\multicolumn{2} {c|}{Singlet} & \multicolumn{2}{c|} {Triplet}\\ \hline
& & &    Exact &MSPQMC& Exact &  MSPQMC \\ \hline
2.0 &$4 \times 2$ & 6 & -8.4059 & -8.4261 & -8.3270 & -8.3725 \\ 
2.0 &$3 \times 4$ & 8 & -15.6745 & -15.7533 & -15.8619 & -15.8533 \\ \hline
4.0 &$4 \times 2$ & 6 & -7.4171 & -7.4318 & -7.2358 & -7.2837 \\ 
4.0 &$3 \times 4$ & 8 & -14.1782 & -14.2698 & -14.3631 & -14.3768 \\ \hline
6.0 &$4 \times 2$ & 6 & -6.7880 & -6.9182 & -6.5556 & -6.8617 \\ 
6.0 &$3 \times 4$ & 8 & -13.2061 & -13.3083 & -13.3422 & -13.4012 \\ \hline
\end{tabular}
\end{center}
\end{table}

\begin{table}
\begin{center}
\caption{ MSPQMC energies of the $3 \times 4$ lattice with 8 electrons, singlet  compared with  
exact calculations for averaging at last (lts), middle (mts) and all (ats)
time slices for $U/t = 1.0$ through 6.0. 
Data for $U/t = 6.0$ is with $\Delta\tau = 0.05$.}
\begin{tabular}{|c|cccc|cccc|}   \hline
\multicolumn{1}{|c|}{$U/t$}&
\multicolumn{4}{c|}{Singlet} &
\multicolumn{4}{c|}{Triplet} \\ \hline
 & Exact & L &    A &    M & Exact  & L   & A   & M   \\  \hline
  1.00 &  -16.7148 & -16.7752 &   -16.7660  &     -16.7742 
  & -16.8425 & -16.8409 &   -16.8267 &     -16.8480 \\
  2.00 &  -15.6745 & -15.7533 &   -15.7568  &     -15.7797 
  & -15.8619 & -15.8533 &   -15.8529 &     -15.8654 \\
  3.00 &  -14.8423 & -14.9073 &   -14.9602  &     -14.9657 
  & -15.0422 & -15.0205 &   -15.0327 &     -15.0558 \\
  4.00 &  -14.1782 & -14.2698 &   -14.3115  &     -14.3150 
  & -14.3631 & -14.3768 &   -14.4047 &     -14.4670 \\
  5.00 &  -13.6427 & -13.7470 &   -13.7872  &     -13.8434
  & -13.8039 & -13.8303 &   -13.8617 &     -13.8696 \\
  6.00 &  -13.2061 & -13.3083 &   -13.3984  &     -13.4432 
  & -13.3422  & -13.4012 &   -13.4695 &     -13.4691 \\ \hline
\end{tabular}
\end{center}
\end{table}

\begin{table}
\begin{center}
\caption{MSPQMC spin and charge correlations of the excited singlet states 
of the $4 \times 2 $ ladder with bond-alternation ($t_{rung}=0.9t$, 
compared with exact calculations for averaging at last (L), all (A) 
and middle (M) time slices for $U/t = 2.0$ and $6.0$. 
Data for $U/t = 6.0$ is with $\Delta\tau = 0.05$.}
\begin{tabular}{|c|c|cccc|cccc|}   \hline
\multicolumn{1}{|c|}{$U/t$}&\multicolumn{1}{c|}{$i,j$}&
\multicolumn{4}{c|}{spin} &
\multicolumn{4}{c|}{charge} \\ \hline
& & Exact & L &    A &    M & Exact  & L   & A   & M   \\  \hline
2.0 &1,1 &  0.5940 & 0.551  & 0.592  & 0.595& 0.9050 & 0.950 & 0.908 & 0.905 \\
&1,2 & -0.2369 & -0.211 & -0.224 & -0.224& 0.4681 & 0.452 & 0.469 & 0.468 \\
&1,5 &  0.0204 & 0.008  & 0.020  & 0.012& 0.5066 & 0.505 & 0.504 & 0.506 \\
&1,6 & -0.0787 & -0.066 & -0.070 & -0.065& 0.5504 & 0.555 & 0.549 & 0.549 \\
&1,7 &  0.0088 & 0.006  &  0.012 & 0.013& 0.5157 & 0.510 & 0.516 & 0.518 \\
 \hline
6.0&1,1 &  0.6873 & 0.618 & 0.682 & 0.689&  0.8127 & 0.882 & 0.818 & 0.811 \\
&1,2 & -0.2775 & -0.238 & -0.198 & -0.250&  0.5056 & 0.476 & 0.495 & 0.502 \\
&1,5 &  0.0549 & 0.003 & 0.002 & 0.011&  0.5149 & 0.511 & 0.522 & 0.517 \\
&1,6 & -0.1002 & -0.072 & -0.029 & -0.073&  0.5406 & 0.548 & 0.545 & 0.546 \\
&1,7 &  0.0136 & 0.022 & 0.010 &0.021&  0.5297 & 0.520 & 0.545 & 0.541 \\ 
\hline
\end{tabular}
\end{center}
\end{table}

\begin{table}
\begin{center}
\caption{ Number of occurances of negative signs for "open-shell"/
excited states in two-dimensions. Sample size is 
$N \times20\times5000$ for $U/t < 4.0$ and 
$N \times 40 \times 8000$ for $U/t \geq 4.0$.}
\begin{tabular}{|c|cc|cc|}   \hline
\multicolumn{1}{|c|}{$U/t$}&
\multicolumn{2}{c|}{$4 \times 2$} &
\multicolumn{2}{c|}{$3 \times 4$} \\ \hline
$U/t$&  Singlet & Triplet & Singlet & Triplet   \\ \hline
  1.00 &1 & 0 & 0 &  0 \\
  2.00 &236 & 211  & 55 &  0 \\
  3.00 &1162 & 1297 & 598  &  258 \\
  4.00 &1980 & 2482 & 2444 &  1439 \\
  5.00 &3321 &3777  & 5676 &  4006 \\
  6.00 &3894 & 4855 & 9818 &  7588 \\ \hline
\end{tabular}
\end{center}
\end{table}

\pagebreak
\clearpage

\begin{center}
{\bf {Figure Captions}}
\end{center}

\begin{description}
{\item{{\bf{Figure 1:}}~~Pair binding energies of two holes for
$4n+2$ and $4n$ Hubbard rings. Open symbols correspond to exact
data and filled symbols to MSPQMC data (squares - ring of 6,
triangles - ring of 14, circles - ring of 12 and diamonds - ring
of 16).}}
{\item{{\bf{Figure 2:}}~~Pair binding energy (triangles) of two holes in the 
$3 \times 4$ lattice with 10 electrons and difference between 
the high and low apin states (squares) of the $3 \times 4$ lattice with
8 electrons. Open symbols correspond to exact data and filled sysmbols 
to MSPQMC data.}}
\end{description}
\end{document}